# Photochemistry upon charge separation in triphenylamine derivatives from fs to µs

Hendrik J. Brockmann[1,4,#], Letao Huang[2,#], Felix Hainer[1,4], Danyellen Galindo[1,4], Angelina Jocic,[3] Milan Kivala[3*], Andreas Dreuw[2,4*], Tiago Buckup[1,4*]

[1]Physikalisch-Chemisches Institut, Ruprecht-Karls Universität Heidelberg, D-69120, Heidelberg, Germany

[2]Interdisziplinäres Zentrum für Wissenschaftliches Rechnen, Ruprecht-Karls Universität Heidelberg, D-69120, Heidelberg, Germany

[3]Organisch-Chemisches Institut, Ruprecht-Karls Universität Heidelberg, Im Neuenheimer Feld 270, D-69120, Heidelberg, Germany

[4]Institute for Molecular Systems Engineering and Advanced Materials, Im Neuenheimer Feld 225, D-69120 Heidelberg, Germany.

*corresponding authors

**Abstract:** Quantum chemical methods and time-resolved laser spectroscopy are employed to elucidate ultrafast charge separation processes in triphenylamine (TPA) derivatives upon photoexcitation. When changing the ambient solvent from generic ones to those capable of accepting electrons, such as chloroform, a vastly extended and multifaceted photochemistry is observed. Following the initial excitation, two concurrent charge transfer processes are identified. Firstly, when the TPA derivative and solvent molecules are correctly positioned, an electron transfer to the solvent molecule with immediate charge separation takes place. Consequently, this process gives rise to the formation of the corresponding radical cation of the TPA derivative. This highly reactive species can subsequently combine with other TPA derivative molecules to yield dimeric species. Secondly, when the molecular positioning upon photoexcitation is not optimal, relaxation back to the $S_1$ state occurs. From this state, an electron transfer process leads to the formation of a charge transfer complex. In this complex, the negatively charged solvent molecule remains closely associated with the positively charged TPA derivative. Within 30 picoseconds, the charges within this complex recombine, yielding a triplet state. This transition to the triplet state is driven by a lower reaction barrier for charge separation compared to the formation of the singlet state.

# 1 INTRODUCTION

Most cutting-edge solar cells being studied at present rely on organic compounds as hole transport materials.[1, 2] Although there is a great variety of hole transport materials, organic semiconductors have emerged as the frontrunners in advancing next generation perovskite solar cell technology.[3] This dominance is particularly noteworthy since the adoption of Spiro-OMeTAD as hole transport material in 2012, which propelled solar cell efficiency to over 9% at that time.[4] Subsequently, the efficacy of triphenylamine (TPA) derivatives for this purpose has been firmly established. However, a significant challenge that persists for organic hole transport materials is their relatively low hole mobility.[5, 6] It becomes thus imperative to conduct comprehensive investigations into the fundamental properties of these materials concerning their photochemical and photophysical attributes.

The past studies of the photochemical behavior of TPA focused primarily on its application in organic synthesis. Since the earliest explorations into TPA's photochemical properties it has been established that upon photoexcitation, TPA tends to undergo a ring-closure reaction, resulting in the formation of the corresponding carbazole compound.[7, 8] This reaction has been extensively scrutinized across a spectrum of polar and non-polar solvents[9-12], with the underlying mechanism claimed to be sequential in nature.[13] Following photoexcitation, the TPA triplet state is efficiently populated through a highly effective intersystem crossing with a quantum yield exceeding 90%.[14] Subsequently, the dihydrocarbazole is formed, initially in a triplet configuration,[11] which then transitions to its singlet ground state as a pivotal intermediate. This intermediate adopts a zwitterionic character[15] and ultimately reverts back to the TPA ground state or undergoes disproportionation into the respective carbazole and tetrahydrocarbazole.[11] It was then found that this reaction is strongly dependent on the concentration of oxygen within the solution. Here, oxygen serves a dual role: on the one hand, it quenches the TPA triplet state, and on the other hand, it enhances the formation of carbazole from the intermediate upon oxidation, thereby augmenting the overall yield.[16] Furthermore, TPA is renowned for its electrochemical reactivity. It has been demonstrated that TPA can facilitate transition metal-mediated single electron transfer reactions,[17, 18] thereby generating *N,N,N',N'*-tetraphenylbenzidine (TPB) from the intermediate TPA radical cation.[19, 20] Given the widespread utilization of TPA derivatives in various emerging technologies, there is a promising avenue for further exploration of their potential in hole formation and transportation. These materials exhibit a propensity to release electrons, making it intriguing to investigate how they perform in electron-accepting environments, offering valuable insights into their behavior under practical conditions.[2]

In this work, we investigate the rapid processes observed upon photoexcitation of TPA in chloroform, which was selected due to its well-known electron-accepting characteristics.[21, 22] To prevent the potential formation of the corresponding carbazole compound, we introduced a second molecule, dimethylmethylene bridged TPA (DTPA, Figure 1). In contrast to TPA's propeller-shaped structure,[23] DTPA features the ortho-carbon atoms of the phenyl rings connected by dimethylmethylene bridges. This structural alteration serves a dual purpose: it not only obstructs the *ortho*-carbon atoms, thus preventing ring-closure, but it also markedly enhances the molecule's rigidity. It forces the molecule into a more planar configuration, leading to greater overlap of the atomic p-orbitals and the extension of the π-system.[23] These structural changes are expected to exert a profound influence on the electronic behavior of the compound. To elucidate the effects of these aforementioned properties, TPA and DTPA are selected as representative examples to investigate the hole formation within this type of molecule. High-resolution broadband transient absorption spectroscopy, covering a time range from femtoseconds to microseconds, in conjunction with quantum chemical calculations, were employed to shed light on this process.

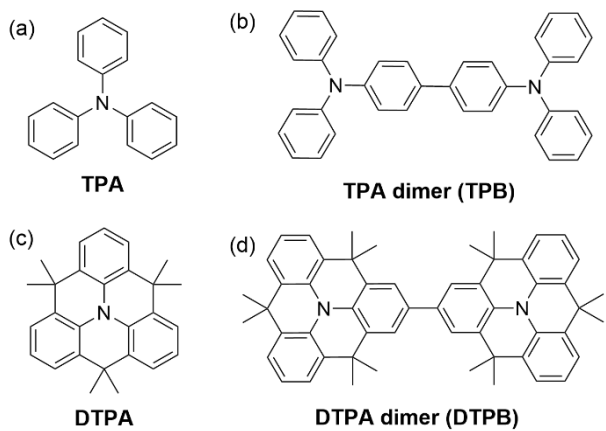

**Figure 1:** a) and b) show the molecular structures of triphenylamine (TPA) and its dimethylmethylene bridged counterpart (DTPA), respectively. c) and d) show the dimeric photoproducts of TPA and DTPA obtained in chloroform, TPB and DTPB

## 2 Experimental and Computational Details

### 2.1 Sample preparation

TPA was purchased from Fisher Scientific. DTPA was synthesized according to literature.[24] Sample solutions for the spectroscopic experiments with optical densities (OD) of 1 at the utilized pump wavelength and cuvette were prepared by dissolving the substances in n-hexane or chloroform.

### 2.2 Spectroscopy

Irradiation experiments under ambient as well as oxygen free conditions were carried out utilizing a 2 mm fused silica cell with a magnetic stirrer and the unfocused pump beam (0.5 mW, 300 nm). The absorption spectra were recorded using a Shimadzu spectrophotometer UV-2600.

Femtosecond transient absorption spectroscopy was used to study the dynamics of TPA and DTPA over 11 orders of magnitude in time, i.e. from the fs to the µs timescale using a Titan:Sapphire laser. The excitation pulse was generated using the commercially available OPA model TOPAS Prime by Light Conversion. It was pumped at 800 nm with a repetition rate of 4 kHz. The generated excitation spectrum was centered around 300 nm with a pulse duration of 90 fs.

For probing, two different light sources were used depending on the timescale investigated. For the longer time scale i.e. from ns up to several µs, a separate broadband fiber-based laser (Leukos STM-2-UV) was used for probe beam generation. The time delay between excitation and probe pulses was scanned electronically in this configuration. For the shorter timescale, ranging from fs up to several ns, the pump-probe delay was scanned mechanically. The white light continuum was thereby generated either in a 1 mm thick sapphire substrate (from 450 to 800 nm) or a calcium fluoride crystal (from 325 to 700 nm). In this configuration the relative polarization of the excitation and probe beam was set to 55° to account for rotational diffusion. The excitation energies for all the experiments were kept constant at 200 nJ. The white-light source for the µs measurements is depolarized. All beams were focused with independent concave mirrors within an angle smaller than 5 degrees. Excitation and probe spot diameter at sample position were about 200 and 100 µm, respectively. All samples were measured in both n-hexane and chloroform. For measurements in n-hexane a 2 mm fused silica cell with a magnetic stirrer was used. For generating the nitrogen atmosphere, the cuvette was equipped with a Normag VH corner globe valve and tightly sealed after removing all oxygen using the freeze-pump-thaw method. Measurements in chloroform, were carried out using a flow-through cuvette. It was operated with a Masterflex pump model from Cole-Parmer. The tubing used for solution transport from the reservoir to the cuvette consisted of chemically resistant PTFE tubes.

### 2.3 Computational Methods

All calculations were performed with Gaussian16.[25] In the following, detailed protocols for the various topics addressed in this work are described.

The geometry optimization and frequency calculations were performed at the B3LYP-D3(BJ)/def2-SVP level of theory, unless stated otherwise. The absorption spectra of both, the ground state and excited states were studied using the PBE0/def2-TZVP level of theory. However, for the charge-transfer chloroform complexes of TPA and DTPA, the absorption spectra were computed utilizing the range-separated exchange-correlation functional CAM-B3LYP to account for the underestimation of charge-transfer excitation energies by most other common exchange-correlation functionals.[26] To take into account the influence of the experimental solvent environment on the geometrical properties, the integral equation formalism polarizable continuum model (IEFPCM) for the solvent (chloroform) was used.[27] [28] All spectra were plotted employing Gaussian functions with constant values for the standard deviation for each molecule.

The hole-electron analysis module provides a very comprehensive characterization of all types of electronic excitations.[29] Three indexes were chosen to characterize the separation of electrons and holes: (1) The *D*-index characterizes the centroid distance of electrons and holes. (2) The *Sr* represents the overlap of electrons and holes. (3) The *t*-index measures whether there is a significant separation of electrons and holes. This information was obtained for the charge transfer chloroform complexes [TPA•Chl] and [DTPA•Chl] employing the Multiwfn 3.8 code.[30]

Based on the results from the electron-hole analysis, the degree of CT of the electronic states of [TPA•Chl] and [DTPA•Chl] could be determined. The geometry and frequencies then were optimized at the CAM-B3LYP/def2SVP level of theory. In order to reliably perform geometry optimizations on a desired excited state potential energy surface using a state tracking algorithms, natural transition orbitals (NTO) analyses were employed at each optimized excited-state geometry, to reassure that it was staying on the desired excited-state potential energy surface.

The total Gibbs free energy has been computes as the sum of electronic energy using the PCM model described above and thermodynamic corrections in gas-phase at 298 K and 1 atm in kcal/mol. Electronic energies were calculated at the M062X/def2TZVPD level of theory. Thermodynamic corrections were obtained from the molecular partition functions using the harmonic oscillator and fixed rotator approximations. The energy barriers for the single electron transfer (SET) processeswere computed by applying the Savéant's "sticky" concerted dissociative electron transfer (cDET) model.[31, 32] The activation barrier (ΔG‡) of the "sticky" cDET can be estimated by:

$$\Delta G^{\ddagger}_{sticky} = \frac{\lambda_{sticky}}{4}\left(1 + \Delta G_0 - \frac{D_P}{\lambda_{sticky}}\right)^2 \qquad (1)$$

with

$$\lambda_{sticky} = \lambda_i + \lambda_0 + \left(\sqrt{D_R} - \sqrt{D_P}\right)^2 \qquad (2)$$

$\lambda_i$ represents the internal reorganization energy, $\lambda_0$ is the external reorganization energy, $D_R$ is the charge-dipole interaction, and $D_P$ is the interaction energy in the corresponding radical-ion pair.

## 3  RESULTS

This section is organized as follows. First, the steady-state absorption spectra of TPA and DTPA and the general observations made by changing the environment are described. Then computed vertical excitation energies of DTPA and the corresponding transient species are given. These allow for the simulation of absorption spectra for the different species. Furthermore, the computation of the Gibbs free energy of potential species should give an insight into possible excited state transitions. Our transient absorption measurements are described at the end.

### 3.1  Steady state Absorption and Irradiation Series

When irradiating TPA and DTPA dissolved in chloroform with 300 nm laser light, the main absorption band at 300 nm decreases and is blue shifted by 10 nm during irradiation (Figure 2). Simultaneously, a new band at 360 nm for TPA and 390 nm for DTPA emerges. This is in stark contrast to the carbazole formation known from literature, which shows its main absorption feature at 293 nm with a shoulder at 340 nm.[11] Instead, the newly generated absorption spectra can be mainly associated with the TPA and DTPA dimers:[33] TPB and DTPB as depicted in Figure 1. This novel dimer-forming reaction pathway remains unaffected by the presence of oxygen within the solution, which also stands in stark contrast to carbazole formation, known from literature.[16] To shed light on this photochemical behavior of TPA derivatives, computational as well as time resolved spectroscopy methods are employed.

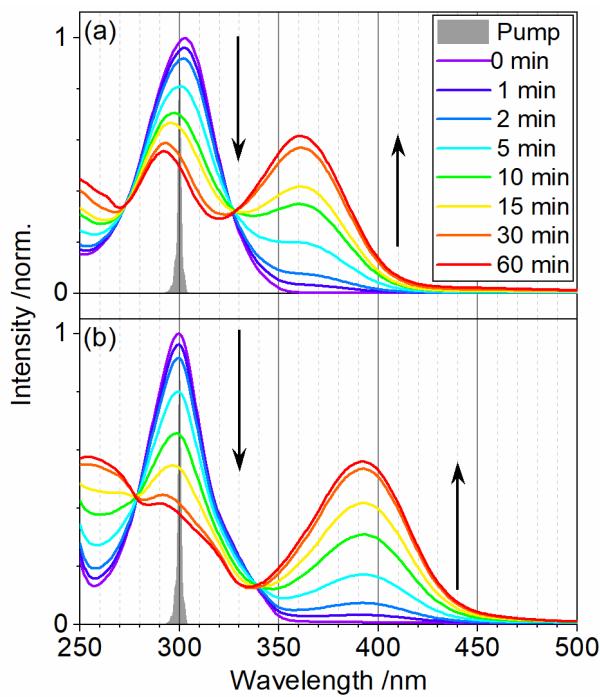

**Figure 2:** Irradiation series of TPA (a) and DTPA (b) in chloroform under ambient conditions. The gray shaded area represents the pump spectrum. The absorption spectra are normalized to the highest maximum with no irradiation.

### 3.2 Quantum Chemical Calculations

Linear-response time-dependent density functional theory[34] (TDDFT) using the PBE0/def2-TZVP xc-functional and basis set combination was employed to calculate the vertical excitation energies and oscillator strengths of the energetically low-lying singlet states of TPA, DTPA and the corresponding dimers TPB and DTPB (Table 1). It is found that TPA possesses a $S_0 \rightarrow S_2$ and $S_0 \rightarrow S_3$ transition at 3.97 eV dominated by the transition from the HOMO to LUMO and HOMO to LUMO+2 (Table 1), which corresponds to the minimum energy peak at 300 nm (4.13 eV) in the experimental steady state UV/VIS absorption spectra in Figure 2. The $S_1$ state of TPA possesses practically no oscillator strength and is thus not visible in the experimental absorption spectrum. The peak evolving at 360 nm (3.44 eV) after irradiation relates to the HOMO-LUMO transition of TPB, which has a computed vertical excitation energy of 3.30 eV (Table 1).

Due to the slightly altered molecular and thus electronic structure of DTPA in comparison to TPA, the vertical excitation energy of the $S_1$ state is 0.4 eV higher with a value of 4.17 eV and the oscillator strength for this transition is increased significantly. In the molecular orbital picture, it is essentially described as electronic HOMO→LUMO transition (90%) and may partially be responsible for the redshifted shoulder in the peak at 300 to 330 nm. Furthermore, the calculated $S_0 \rightarrow S_3$ and $S_0 \rightarrow S_4$ transition exhibit an excitation energy of 4.29 eV and an oscillator strength of 0.16, which is mainly responsible for the slightly blue shifted main absorption band of DTPA in comparison to TPA. After irradiation, this peak shifts to 390 nm (3.18 eV) which correlates to the $S_0 \rightarrow S_1$ transition of DTPB, the $S_1$ state of which can be characterized as a single-electron HOMO→LUMO transition. Furthermore, visual inspection of the frontier molecular orbitals (Figure SI 2) reveals that the extension of the π-system influences mostly the LUMO, LUMO+1, LUMO+3 and not so much the HOMO and LUMO+2, however, they change their energetic order when going from TPA to DTPA.

**Table 1:** Vertical singlet excitation energies of TPA, DTPA and their dimers TPB and DTPB in eV and the corresponding oscillator strengths (brackets) computed at the TDDFT/PBE0/def2-TZVP level of theory.

|  | TPA | TPB | DTPA | DTPB |
|---|---|---|---|---|
| $S_1$ | 3.77 (0.02) | 3.30 (1.28) | 4.17 (0.16) | 3.35 (0.96) |
| $S_2$ | 3.97 (0.32) | 3.62 (0.02) | 4.17 (0.16) | 3.70 (0.02) |
| $S_3$ | 3.97 (0.32) | 3.69 (0.00) | 4.29 (0.15) | 3.86 (0.16) |
| S4 | 4.48 (0.05) | 3.81 (0.00) | 4.29 (0.15) | 4.00 (0.09) |
| S5 | 4.48 (0.05) | 3.87 (0.08) | 4.75 (0.00) | 4.01 (0.10) |

Furthermore, the steady-state absorption spectra for potential transient species occurring during the transient absorption experiments have been simulated. The vertical excited-state absorption spectrum of the first excited singlet state of DTPA has been calculated at its independently optimized equilibrium geometry. The main feature is located at 1.83 eV (679 nm) with a significant shoulder at 2.05 eV (604 nm) (Figure 3). Furthermore, two less pronounced peaks at 3.66 eV (339 nm) and 3.48 eV (356 nm) are obtained at the theoretical level of TDDFT/PBE0/def2-TZVP employing the PCM model for chloroform solvation. The corresponding excited-state absorption spectrum of the first triplet excited state exhibits a broad peak at around 700 nm, which consist of two electronic transitions with excitation energies of 1.89 eV (657 nm) and 1.74 eV (711 nm), respectively. Next to this, a very small peak at 2.27 eV (547 nm) can be observed. Nevertheless, a main feature is located at 2.70 eV (460 nm) with a shoulder caused by transitions at 2.90 eV (427 nm) and 2.94 eV (421 nm). Also, the absorption spectrum of the DTPA radical cation is calculated for comparison. Its most pronounced peak is found at 580 nm, caused by in fact two electronic transitions at 2.18 eV (569 nm) and 2.13 eV (581 nm). Also, a small absorption band at 3.53 eV (351 nm) is seen in the computed spectrum of the DTPA radical cation. The corresponding excited-state spectra of the first singlet and triplet states of TPA, and the absorption spectrum of the TPA radical cation are presented in the SI.

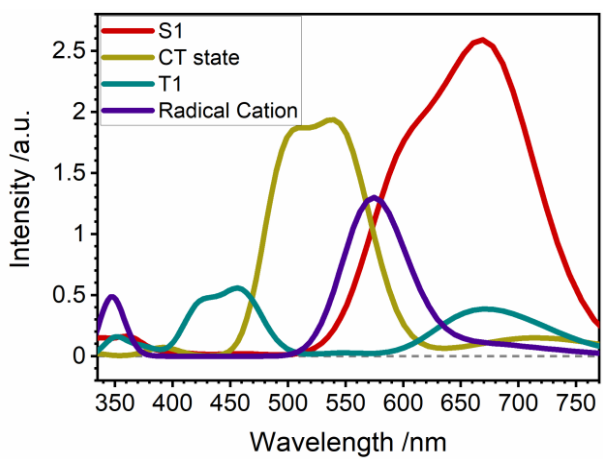

**Figure 3:** Simulated transient absorption spectra for potential short-lived DTPA species. The spectra have been generated by convoluting the calculated stick spectra (energies and oscillator strengths) with Gaussian broadening functions with a standard deviation of 0.24 eV.

Previously Fitzgerald et al. made the observation that amines similar to the ones investigated in this work could, upon photoexcitation, undergo single electron transfers when dissolved in chloroform.[35] Therefore, also possible charge-transfer complexes have been investigated theoretically. The geometries of [TPA•Chl] and [DTPA•Chl] complexes have been computed and their vertical excited states calculated employing DFT and TDDFT using now CAM-B3LYP as xc-functional (see above) and the PCM model for chloroform. The employed electron and hole analyses of the excited electronic states indeed reveals their substantial charge-transfer character enabling electron transfer following the excitation process. As can be seen from the data in Table SI 1, the sixth excited singlet state $S_6$ of [DTPA•Chl] has a charge transfer index (D) of 4.08 Å and an electron-hole separation of 2.51 Å, while the overlap of hole and electron ($S_r$) reaches only 0.20 a.u. Similar results are found for the sixth excited singlet state $S_6$ of [TPA•Chl], the D index is as high as 3.50 Å, the t-index reaches 2.29 Å and the overlap $S_r$ is only 0.30 a.u. (Table SI 2). The detailed nature of this electron transfer state is visualized in Figure 4. In detail, the hole is mostly located at the nitrogen atom in DTPA or TPA with 17.95 % and 22.42%, respectively, while the electron is large localized at the C atom of chloroform with 12.12 % and 24.21% in [DTPA•Chl] and [TPA•Chl], respectively.

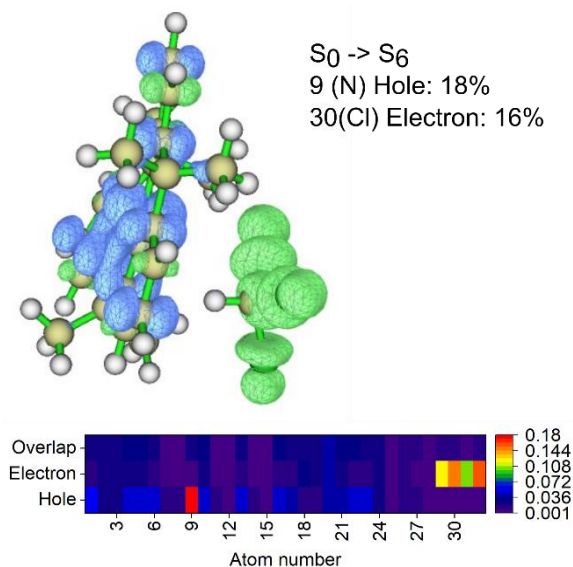

**Figure 4:** The electron (green) and hole (blue) distribution for the DTPA-chloroform charge transfer complex (DTPAchl). The atomic number refers to the first 32 non-hydrogen atoms, where 9 is the nitrogen atom of DTPA, 29 the carbon and 30 to 32 are the chlorine atoms of chloroform.

Although this electron transfer state is only the sixth lowest excited singlet state $S_6$ at the Franck-Condon (FC) geometry, i.e. the equilibrium geometry of the electronic ground state of [DTPA•Chl] it quickly becomes the lowest excited $S_1$ state when its geometry is optimized. To make sure the geometry optimization follows the correct potential energy surface a state-tracking algorithm has been employed, which is supported by the results of the NTO analysis. The energy of the equilibrium geometry of the CT state is found to be 12.06 kcal/mol lower than that of the $S_1$ state. The corresponding triplet charge transfer state is found to be even lower by another 10.26 kcal/mol. To compare with the transient absorption measurements below, the excited state absorption spectrum of the CT state of the [DTPA•Chl] complex is calculated (Figure 3). At the TDDFT/CAM-B3LYP/def2-TZVP level, the main transition is given at 2.27 eV (547 nm) with a pronounced shoulder at 2.48 eV (499 nm). An additional broad band is located at 1.78 eV (698 nm), accompanied by several smaller features at 3.15 eV (394 nm) and 4.12 eV (301 nm) along with a distinct band at 5.23 eV (237 nm) (Figure 3).

At the equilibrium structure of the CT state of the [DTPA•Chl] complex, the transferred electron to the chloroform molecule induces a significant elongation of one of the carbon-chlorine single bonds to 2.64 Angstroms as compared to about 1.74 Å of a regular C-Cl bond. Therefore, the dissociation of the [DTPA•Chl] complex into a DTPA radical cation, a $CHCl_2$ radical and a chloride anion is proposed. The "sticky" concerted dissociative electron transfer model is employed to check the feasibility for this process with DTPA and TPA. The obtained energy barriers to form the TPA or DTPA radical cations from the [TPA•Chl] and [DTPA•Chl] charge transfer states amount to 8.58 kcal/mol and 9.04 kcal/mol, respectively. The Gibbs free energy change of the whole radical formation processes described above are -36.24 kcal/mol and -32.74 kcal/mol, respectively. Eventually, the radical cations can then form the final photoproducts, i.e. DTPB and TPB, with a Gibbs free energy release of -39.74 kcal/mol and -43.96 kcal/mol.

### 3.3   Femtosecond Transient Absorption Measurements

Both TPA and DTPA are dissolved in chloroform and time resolved transient absorption is measured. The overall shape and evolution of the transient absorption data for these molecules exhibit remarkable similarity. Consequently, in this discussion, we will concentrate on the data associated with DTPA, although it is worth noting that the data for TPA underwent the same treatment, as demonstrated in the Supplementary Information (SI).

When examining the 2D dataset for the timescale up to 8 ns delay time in Figure 5a, a multifaceted picture unfolds. During this early timeframe, the samples display rapid dynamics. At earliest time, the signal has a maximum at 580 nm. Another distinct ESA band is visible at 355 nm, characterized by its narrow spectral profile spanning just 20 nm. These bands exhibit decay throughout the entire temporal measurement range.

However, it's important to note that this decay doesn't occur uniformly across all wavelengths. Consequently, around the maximum at 580 nm the blue shifted flank decays slower than the red shifted one. This observation suggests a spectral overlap involving multiple species. At a time delay of approximately 20 ps, another ESA Band emerges. This band's spectral maximum is situated at 420 nm. In the 2D dataset encompassing longer delay times of up to 450 microseconds, it becomes evident that the initially described ESA band at 580 nm does not completely vanish.

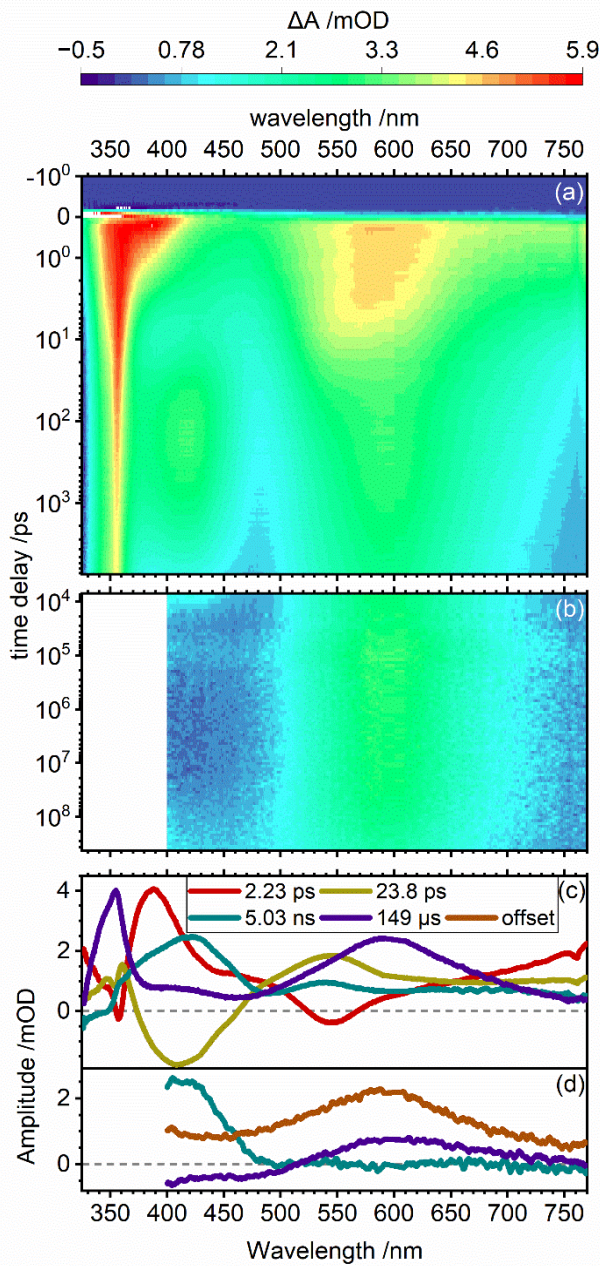

**Figure 5:** Two-dimensional representation of transient absorption data of DTPA in chloroform for the short delay experiment (top) and the long delay experiment (middle). Bottom: DADS obtained from simultaneous global analysis of both datasets with the upper graph showing the results using the short delay data and the bottom graph showing the results using the long delay data.

For global analysis the data sets of all measurements are combined and simultaneously fitted. A four exponential function is used for the fit to assign explicit time constants to the decays. It should be noted that the fits are performed with an additional Heaviside function as an offset to account for the still clearly visible signal even at the maximum delay time of 450 µs. This is to describe the apparent formation of a photo product and is therefore essential for sufficient fitting of the data. However, the last observed species seems to decay very slowly, while some positive signal at the short wavelength edge of the spectral range emerges. The resulting DADS are shown in Figures 5c) and 5d). The DADS for the initially observed decay with a time constant of 2.23 ps show a sharp positive signal at 390 nm and a very broad band extending from around 600 nm all the way to the long-wavelength edge of the spectrum. In this first DADS, the formation of the second species is indicated by a slightly negative signal at 540 nm for the first decay. This signal coincides with the maximum observed in the second DADS with a decay constant of 23.8 ps, indicating a sequential process. The same applies to the combination of minimum and maximum at 420 nm. The second DADS also has negative contributions at 420 nm, which correspond to the formed positive band in the third DADS. This last DADS has a time constant of 5.03 ns and does not show any negative contribution itself. Only at the ultraviolet edge of the recorded spectrum is there a tendency towards negative values, which can be attributed to the decay of the ground state bleaching. The remaining fourth DADS with a decay constant of 149 µs exhibits maximums at 600 and 355 nm.

## 4   DISCUSSION

During the forthcoming discussion, we will initially explore the viability of excited-state charge transfer transitions for TPA derivatives in solvents with electron-accepting properties. Subsequently, we will delve into a comprehensive examination of the experimental transient absorption signal. The hereby developed kinetic model provides the necessary information to model the time-dependent concentrations of all involved species from fs to µs and enables us to compare the related rate constants and species spectra to the results from the quantum chemical calculations.

The calculated activation energy for the generation of radical cations from DTPA or TPA with the corresponding chloroform anion is 8.58 kcal/mol or 9.04 kcal/mol, respectively. In general, the process is strongly exergonic with a total Gibbs Free energy of -36.24 kcal/mol or -32.74 kcal/mol. This shows that the single electron transfer from the respective TPA derivative to the solvent is feasible. The concluding formation of the potential photoproducts TPB and DTPA with total Free gibs energies of -39.74 kcal/mol and -43.96 kcal/mol also is energetically viable.

Furthermore, the electron-hole analysis shows that there are charge transfer states present for TPA as well as DTPA. Even though they are energetically higher lying at Franck-Condon conditions, we were able to show that they get accessible from relaxed excited state geometries. Next to this, it was shown that the free energy gain for the CT state to form the $T_1$ state is much greater than for the generation of the $S_1$ state. This indicates the formation of the triplet excited state after the formation of the CT state from the singlet state.

This makes clear that not only the formation of DTPB from DTPA is energetically possible, but also involves single electron transfers and the formation of various chare transfer states. When trying to unravel the underlying mechanism behind the formation of these states and subsequently the formation of the observed photoproduct, the first thing to consider is the longest-lived decay in the DADS described in the results section above. This DADS with a decay constant of 149 µs does not seem to be formed from any of the other species. Therefore, it does not evolve from the same sequential process as the other three. In particular, the presented spectral overlap of different ESA bands across various segments of the delay-time axis provides compelling evidence of this phenomenon. Such behavior defies explanation within a sequential model, as it necessitates the simultaneous coexistence of distinct species with differing lifespans. This is noteworthy within two respects. Firstly, other than the remaining species this last one seems to not decay back to the ground state but form the final photoproduct. Secondly, this stands in bright contrast to all previously studied cases of photoexcitation-based reaction on TPA derivatives in non-electron accepting solvents, as they usually follow a clear sequential relaxation process towards the observed photoproduct.

Consequently, it is essential to introduce a branching process into the kinetic model. The transient absorption data clearly signifies the presence of at least two distinct species right from the inception of the resolvable time scale. Therefore, this branching must be occurring prior to the discernible time delays at hand. We infer that the efficiency of electron transfer and, consequently, dimer generation is contingent upon the initial arrangement of donor and acceptor molecules immediately after photoexcitation. Thus, depending on the solvent environment, two distinct relaxation pathways become viable. In one scenario, an immediate electron transfer ensues, leading to the productive reaction pathway. In an alternate scenario, the charge transfer occurs in a delayed fashion, prompting the adoption of the unproductive pathway back to the ground state.

In this second mentioned case, cooling from the initially populated hot $S_{1/2}$ to the S1 vibrational ground state occurs first. Given the lower energy of this state, the charge transfer to the solvent remains feasible; however, a complete charge separation does not occur due to the molecules lacking the necessary energy to surmount the emerging Coulomb barrier. Thereby the charges remain close to each other and are therefore able to recombine, generating the more stable triplet ground state, subsequently repopulating the singlet ground state.

Based on these evaluations, the kinetic model showcased in Figure 6 can be deduced, subsequently resulting in the development of the SADS depicted in Figure 7. In this specific context, the initial branching factors were configured at a 1:1 ratio. However, it is important to note that in this instance, these factors signify starting populations rather than true branching ratios. While it remains challenging to precisely determine the actual values for these starting ratios, their precision does not impact the qualitative analysis of the spectra or the quantitative assessment of the time constants.

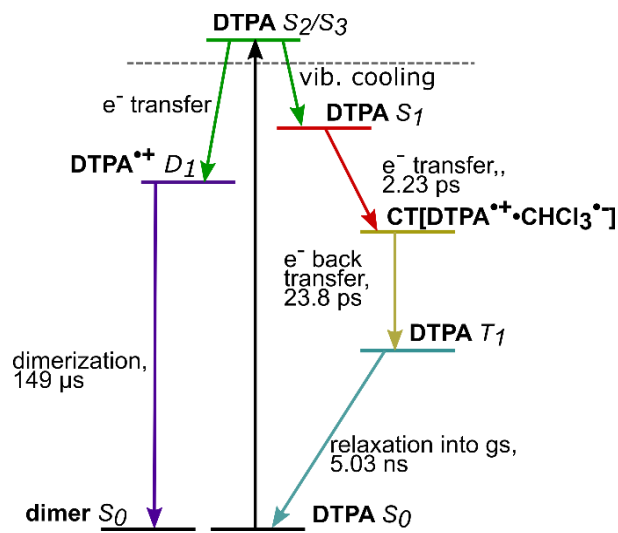

**Figure 6:** Kinetic model summarizing the photophysics and photochemistry DTPA in chloroform.

The SADS (Species associated difference spectra) shown in Figure 7a) for the initial observed state of DTPA in chloroform with a time constant of 2.23 ps displays significant resemblances to SADS for the DTPA $S_1$ state from comparable measurements in nonpolar solvents (See SI). The maximum occurs in the infrared wavelengths area outside of the measured spectral space. Additionally, there is another band at 370 nm. In contrast, the subsequent species with a decay constant of 23.8 ps is notably blue shifted compared to this first species, with a peak at 545 nm. The band in the ultraviolet region of the spectrum is also blue shifted by approximately 20 nm and is now located at 355 nm. This spectrum we assign to the lower energy charge transfer complex.

The final species, which we assign the triplet to, which relaxes back to the ground state through the relaxation pathway labeled "unproductive" due to the lack of discernible evidence of subsequent species, relaxes with a time constant of 5.03 ns. Its peak in the transient absorption (TA) spectrum is at 420 nm. In this representation, the peak of the longest-lived species from the productive reaction pathway is well observed at 600 nm, with a lifetime of 149 µs. This intermediate species also displays a pronounced secondary peak at 350 nm. This secondary peak is quite sharp, spanning about 20 nm. The long live time of this species can be well explained by very low concentrations in the sample solution, as intermolecular reaction steps are significantly affected, as they would be expected to follow the formation of the radical cation in forming the dimer as the expected photoproduct. This assumption is further supported by comparing this spectrum for TPA (SI) with literature reported UV/VIS spectra. All these spectra fit very well with the ones obtained by quantum mechanical calculations.

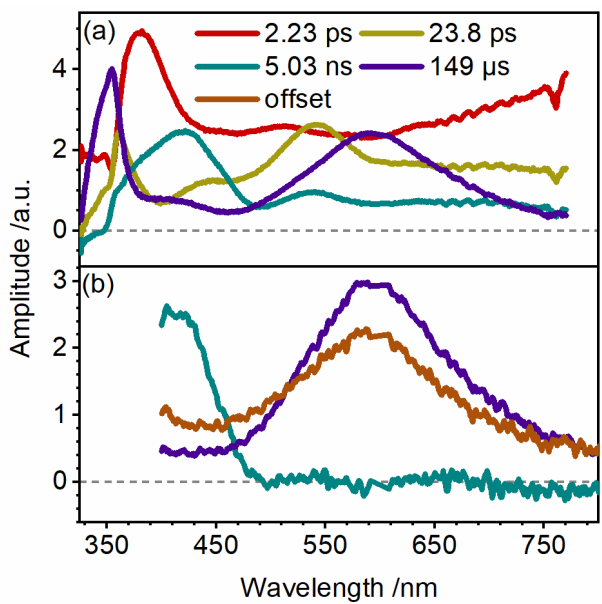

**Figure 7:** SADS obtained for parallel decay of DTPA $S_1$ and DTPA•+ in the TA-Data of DTPA in chloroform with a) SADS from short delay data and b) SADS from long delay data.

Lastly, an increase in amplitude at the ultraviolet edge of the spectrum can be observed, suggesting the onset of formation of a subsequent species localized in that region, which can be well observed for wavelengths below 500 nm. Comparing the offset to the steady state absorption spectrum from figure 2 shows that this cannot be the final photoproduct.

## 5  CONCLUSIONS

Charge transfer states and charge separation processes in TPA derivatives were characterized by quantum chemical calculations and femtosecond transient absorption spectroscopy. In this work the environment of the molecules was changed to electron accepting conditions which opened up not yet observed reaction pathways upon photoexcitation. Instead of the previously reported formation of the respective carbazoles the formation of dimer species was observed. If a solvent molecule was positioned correctly towards the TPA derivative an electron transfer reaction with prompt charge separation happened, forming the corresponding radical cation. This species subsequently reacted with other TPA derivative molecules to form the dimer species. Alternatively, if solvent and TPA derivative molecules are positioned in a less ideal way, first relaxation to the $S_1$ state occurred. From this state then, an electron transfer within a few picoseconds was observed. In this case though, the solvent anion remained close to the positively charged TPA derivative. Consequently, after around 30 picoseconds charge recombination occurs yielding the TPA derivative in its triplet state.

A detailed model for these processes was developed which allowed identification of the involved species and conformation of those by comparing calculated spectra to practical measured ones. Furthermore, the reaction pathway for the generation of DTPB and TPB was verified by DFT calculations of the free Gibbs energy for the various processes.

Summarizing, this work reveals a competitive process for charge transfer processes in the investigated TPA derivatives. It was found that both possible charge transfer reactions outpace and therefore inhibit the previously observed carbazole formation. Furthermore, it was found that the efficiency of the charge separation after the electron transfer reaction is determined already within in the moment of excitation and not due to further reorientation of the corresponding molecules. These insights on charge transfer and separation processes and the balancing between the now three known possible photochemical pathways could be an interesting guideline for the design aspect of TPA-based materials.

**ASSOCIATED CONTENT**

**Supporting Information**. 1. Calculated spectra for TPA, DTPA and their transient species. 2. Computational details on the electron-hole analysis. 3. Computational details on the single electron transfer and NTO analysis. 4. Dynamics for TPA. 5. Dynamics of DTPA and TPA in hexane. 6. Experimental time resolution. 7. Coordinates of the computationally obtained structures of transient species of TPA and DTPA.

**AUTHOR INFORMATION**

Corresponding Authors

tiago.buckup@pci.uni-heidelberg.de

+ 49-6221-548723

milan.kivala@oci.uni-heidelberg.de

+49-6221-5419823

dreuw@uni-heidelberg.de

+49-6221-5414735

Author Contributions

#These authors have contributed equally to this work.

All authors have given approval to the final version of the manuscript.

**ACKNOWLEDGMENT**

All authors thank the Deutsche Forschungsgemeinschaft (DFG) for funding through the Sonderforschungsbereich SFB 1249, TP A05, B01 and B04. Letao Huang acknowledges the Chinese Scholarship Council, as well as Dr. Jie Han for providing useful suggestions.